\documentclass[prl,twocolumn,aps,showpacs,superscriptaddress]{revtex4}
\usepackage{graphics}
\usepackage{epsfig}
\usepackage{amsmath}
\usepackage{amssymb}
\usepackage[usenames]{color}

\begin{document}
\newcommand{\beq}{\begin{equation}}
\newcommand{\eeq}{\end{equation}}
\newcommand{\eps}{\varepsilon}
\newcommand{\Pone}{P_{\chi_1}}
\newcommand{\Ptwo}{P_{\chi_2}}

\title{Shock Waves in Weakly Compressed Granular Media}

\author{Siet van den Wildenberg}
\affiliation{Kamerling Onnes Lab, Universiteit Leiden, Postbus
9504, 2300 RA Leiden, The Netherlands}
\author{Rogier van Loo}
\affiliation{Kamerling Onnes Lab, Universiteit Leiden, Postbus
9504, 2300 RA Leiden, The Netherlands}
\author{Martin van Hecke}
\affiliation{Kamerling Onnes Lab, Universiteit Leiden, Postbus
9504, 2300 RA Leiden, The Netherlands}

\date{\today}

\begin{abstract}
We experimentally probe nonlinear wave propagation
in weakly compressed granular media, and
observe a crossover from quasi-linear sound waves at low impact, to shock waves at high impact. We show that this crossover grows with
the confining pressure $P_0$, whereas the shock wave speed is independent of
$P_0$ --- two hallmarks of granular shocks predicted recently \cite{gomez}. The shocks exhibit powerlaw attenuation, which we model with a  logarithmic law implying that local dissipation is weak. We show that elastic and
potential energy balance in the leading part of the shocks.
\end{abstract}

\maketitle

Many disordered materials, including granular media
\cite{gomez,bobPRL,Cheng,bobfragile},
foams \cite{gijsjam} and emulsions \cite{emulsion,brujic}, lose their rigidity when their confining
pressure $P_0$ is lowered. In almost all cases,
the resulting unjamming transition goes hand in hand with the vanishing of
one or both elastic moduli
\cite{Bolton,Durian,makse,OHernetal,ellenbroek,SS,finsize,review1,review2,footnote1}, and
consequently, nonlinearities must dominate
when such marginal systems
are subjected to finite stresses
\cite{gomez,N&V,ohern}.
For example, soft particles exhibit nonlinear rheology near jamming \cite{emulsion,gijsflow}, even when their local elastic and viscous interactions
are linear \cite{olson,brian}, and marginally connected spring networks exhibit nonlinear elasticity near their critical points
\cite{MW,Fred}. In these two cases, the vanishing of the elastic moduli is a collective
phenomenon, closely connected to the isostatic character of the marginal state \cite{Bolton,Durian,makse,OHernetal,ellenbroek,SS,finsize,review1,review2,footnote1}.

Here we experimentally probe a different scenario
where {\em local} nonlinearities near unjamming lead to
vanishing elastic moduli and nonlinear excitations: shock waves
in granular media \cite{shocknote}.
Granular media have frictional interactions, and these prevent granular media to reach the isostatic limit. Consequently, there are no {\em collective} mechanisms leading to
vanishing elastic moduli or nonlinearities \cite{review1,ellakcritical,Kostya,silke}.
Nevertheless, when granular media unjam as the pressure $P_0$ is lowered, the
individual contacts weaken due to the nonlinear {\em local} Hertz contact law, which states that for elastic
spheres, the contact forces $f$ scale with deformations
$\delta$ as $f \sim \delta^{3/2}$ \cite{Johnson,footnote2}.
As a result, frictional granular media
have a vanishing linear response at their unjamming point, and
their elastic moduli and sound speed vanish as $P_0^{1/3}$
and $P_0^{1/6}$, respectively
\cite{makse,ellaksoundwaves,nesterenko,weakening,jia}.

\begin{figure}[tb]
\includegraphics[width=1\linewidth,clip,viewport=10 320 410 526]{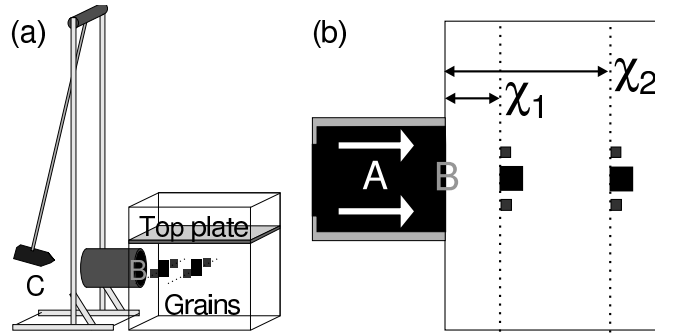}
\caption{Schematic side view (a) and top view (b) of our setup. Shock waves are excited in the granular medium by
a plunger (A) which slides through a circular hole (B) and is impacted by a heavy mass (C). Pressure sensors and accelerometers (big resp. small
squares - both enlarged for visibility) are connected to steel wires immersed in the granular medium
at locations $\chi_1$ and $\chi_2$.}
\end{figure}

Recent simulations on frictionless Hertzian media by Gomez {\em et
al.} suggest that sound waves give way to strongly nonlinear shock
waves near the unjamming
($P_0 \rightarrow 0$) point \cite{gomez}. Three crucial questions remain open, as the numerical model of Gomez has no static friction, no dissipation
and is in 2D. First, realistic granular media are frictional: do shock waves also arise for non-isostatic, frictional systems? Second, friction also leads
to dissipation --- do shock waves survive realistic levels of
dissipation? Third, can such shock waves be excited in 3D
experiments?

To answer these questions, we experimentally probe sound and shock
waves by impacting a weakly compressed granular medium with a heavy
mass, while measuring the propagation speed and front shape for a
wide range of impact magnitudes (Fig.~1). We find a novel power law attenuation of
the shock waves, which we can model with a simple model where local dissipation depends logarithmically on grain forces. Notwithstanding this weak dissipation,
our shock waves exhibit the three main hallmarks
of the numerically observed conservative shocks: {\em (i)} a crossover from linear waves
to shock waves when the impact pressure $P$ exceeds the confining pressure $P_0$; {\em(ii)} independence of the shock speed on $P_0$ but
powerlaw scaling with impact strength; {\em(iii)} a balance of kinetic and potential energies in the leading edge of the shock waves. The ease with which we can excite such granular shock waves suggests that they play an important role whenever loose granular
media are strongly excited \cite{Thoroddsen,detlef,johnjet,ChengSheet,plow}.

\begin{figure}[tb]
\includegraphics[clip,viewport=-20 -20 450 220,width=1\linewidth]{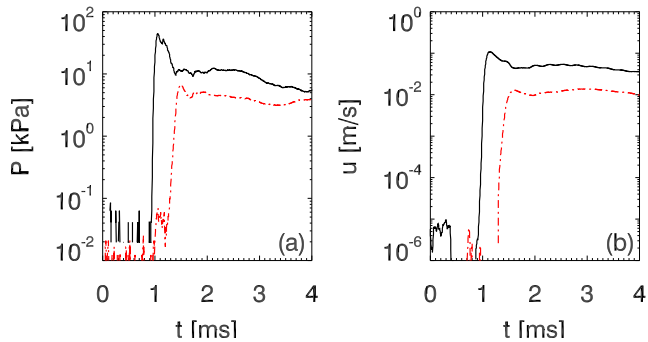}
\caption{(Color online) Typical pressure and velocity signals for
high impact amplitude, for $P_0=0.9$ kPa, at $\chi_1=5$ cm (black)
and $\chi_2=15$ cm (red/dashed). (a) Local pressures $P(t)$. (b)
Local velocities $u(t)$ obtained by integrating the acceleration
signal.}
\end{figure}

{\em Setup ---} Our setup consist of a
large metal container ($45 \times 45 \times 45 $ cm) filled with
glass beads (diameter $D= 3.8-4.4$ mm)
and covered by an 2 kg aluminum top plate. To excite
shock waves, we impact a freely sliding cylindrical piston (A),
(diameter 10 cm, length 18 cm, mass 3.7 kg), which makes contact
with the grains through a circular hole in the side of the
container (B), with a mass (C, 1.2 kg) suspended from a pendulum
(mass 2 kg, length 1.3 m). We detect wave propagation throughout the material
via pressure sensors and
accelerometers burried in the granular medium; the sensors are attached
to steel wires to ensure their
correct positioning and orientation. We  thus
probe the
local excess pressure (our pressure sensors have no DC response)
and acceleration at distances $\chi_1$ and $\chi_2$
from the impact zone (Fig.~1). 
See Supplemental Material at [URL will be inserted by publisher] for experimental details.

{\em Phenomenology ---} In Fig.~2 we show typical time traces of the local pressures and velocities detected at locations
$\chi_1=5$ cm and $\chi_2=15$ cm, for a strong impact and low confining pressure.
The waves take the form of fronts with a clearly identifiable leading edge
where the pressures and particle velocities peak, followed
by a long tail with a complex structure. Here we will focus on the nonlinear regime, where the propagation is set by the amplitude. We characterize our waves by the peak pressures, $\Pone$
and $\Ptwo$, and peak particle velocities, $u_1$ and $u_2$, at $\chi_1$ and
$\chi_2$. We determine the front speed $V_s$ from the time of travel $\Delta t$, where $\Delta t$ is the interval between $P$ reaching its
50\% value at $\chi_1$ and $\chi_2$.

Before embarking on a systematic study of the propagation of these fronts,
we briefly discuss the typical scales shown in Fig.~2.
First, the typical grain velocities are of the order of cm/s, the impact speeds of the plunger are of the order of m/s whereas
$V_s$ is of the order of hundreds of m/s.
Second, the grain displacements at $\chi_1$ at the time of the pressure peak are of the order of 10 $\mu$m, consistent with purely elastic deformations (See Supplemental Material at [URL will be inserted by publisher] for details), whereas the total motion of the plunger into the sand for such impacts is of the order of a mm.

The picture that emerges is that upon impact, a rapid front is formed, which,
for the cases when $V_s$ is larger than the sound speed, we will call a shock wave. We note that we do not observe the breaking up of our fronts in several distinct solitons --- in the simulations of \cite{gomez}, the stability of the shocks was attributed to the disorder of the granular packings. The shock
being nonlinear, its speed is set by its amplitude, and
the leading edge of the shock remains leading, as it outruns whatever happens in the tail. The interactions in the leading edge are dominated by elastic, Hertzian interactions. However, long after the shock wave has outrun our
system, the plunger is still slowly penetrating the sand bed, leading to a very long tail, where the vast majority of rearrangements and dissipative events take place.

\begin{figure}[tb]
\includegraphics[width=1\linewidth,clip,viewport=0 20 410 325]{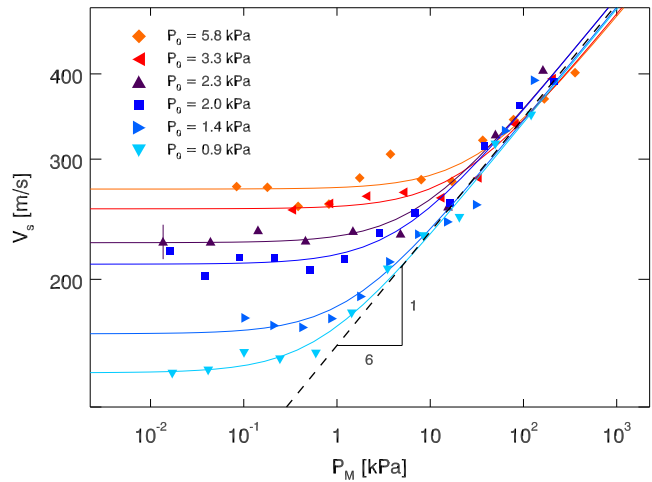}
\caption{(Color online) Front speed $v_s$ as a function of peak
pressure in the front, $P_m$, for a range of confining pressure
$P_0$, at $\chi_1=5$ cm at $\chi_2=15$ cm. Each data point corresponds to
a number of runs, and we indicated a single typical 5\% error bar. Fits (solid lines) are functions of the form $V_s= C(1+P_m/P_f)^{1/6}$. For $P_m \gg P_0$, $V_s \sim P_m^{1/6}$ (dashed line) and is
independent of $P_0$.}
\end{figure}

{\em Propagation speed ---} To probe the nature of the waves excited in this system,
we varied the pressure $P_0$ at the depth of the sensors from 0.9 to 6 kPa and
have determined the propagation speed $V_s$ for a wide range of impact strengths.
To compare our data to the theoretical predictions of \cite{gomez}, we
plot our data in a so-called Hugoniot plot.
Due to attenuation (see Fig.~2), $\Ptwo < \Pone$ and as a measure of the impact
strength we use their geometric mean, $P_m:= \sqrt{\Pone \Ptwo}$.

Fig.~3 shows $V_s$
as a function of $P_{m}$ and $P_0$, and that
our data is well fit by  the form $V_s=C
(1+P_m/P_f)^{1/6}$, which is very
close to the more involved analytical formula for shock waves presented in
\cite{gomez}.

The main three features of our data are {\em (i)} For strong impacts, $V_s$ becomes independent of $P_0$, and
$V_s$ scales consistently as
$P_{m}^{1/6}$ --- the dependence on $P_m$ is a hallmark of nonlinear waves, and the exponent $1/6$ is the one predicted for
Hertzian shock waves. {\em (ii)} For weak impacts, $V_s$ becomes independent of the impact
strength --- the hallmark of linear waves --- but increases with pressure
due to the nonlinear local interaction law.  {\em (iii)} The crossover from the linear to nonlinear regime is expected to arise when
$P_m \gg P_0$, because when $P_m \ll P_0$, the Hertzian interaction can be linearized, whereas for $P_m \gg P_0$,
linearization fails and shocks arise \cite{gomez}. Indeed
we find that $P_f$ grows with
$P_0$.

All these features are in good qualitative and quantitative agreement with the earlier numerical findings of Gomez {\em et. al}: the waves we excite for large impacts are indeed shock waves.

{\em Attenuation ---}  The peak pressure diminishes whilst the shock propagates through the material. In Fig.~4 we compare the
peak pressures $\Pone$ and $\Ptwo$ at $\chi_1=5$ cm and $\chi_2=15 $ cm for the same experiments as shown in Hugoniot plot. Strikingly,
the attenuation varies significantly with impact strength:
for the weakest impacts, $\Ptwo \approx \Pone$, while for larger impacts the
relation between $\Ptwo$ and $\Pone$ is consistent with power law
scaling:
\begin{eqnarray}\label{pwrlw}
\mbox{For } \tilde{P} \ge 1:&  \tilde{P}_2 &= \tilde{P}_1^\beta~, \label{pwrlw1} \\
\mbox{For } \tilde{P} \le 1:&  \tilde{P}_2 &= \tilde{P}_1 \label{pwrlw2},
\end{eqnarray}
where $\tilde{P}:=P/P^*$, $\tilde{P}_{1,2}:=P_{\chi_{1,2}}/P^*$, the characteristic
pressure $P^* \approx 50$ Pa and $\beta \approx 0.77 \pm 0.05$.

\begin{figure}[tb]
\includegraphics[width=1\linewidth,clip,viewport=-20 10 430 337]{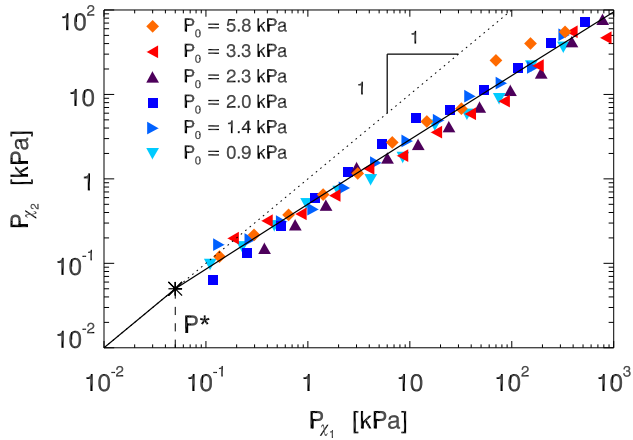}
\caption{(Color online) Scatter plot of the maximum pressures
$\Pone$ and $\Ptwo$ at $x_1=5$ cm and $x_2=15$ cm for a range of
pressures --- same data as in Fig.~3. The black curve is a fit to our
model (Eq.~(\ref{disc})-(\ref{eps2}), See Supplemental Material at [URL will be inserted by publisher] for details.
)}
\end{figure}

We will now determine a simple local model that captures this remarkable powerlaw attenuation. We ignore disorder and imagine the signal to propagate over a linear chain of beads, where each bead attenuates the signal by a factor $(1-\varepsilon)$, where the local attenuation
$\eps$ depends on the local pressure $P$.
After some algebra, we find that we can capture the power law relation between
$\Pone$ and $\Ptwo$ when $\eps(P)$ has the following logarithmic form (See Fig.~5a):
\begin{eqnarray}
\mbox{For } P \ge P_*:&  \eps(P) &= \eps_s ~ ln \left(\frac{P}{P_*}\right)~,\label{eps1}\\
\mbox{For } P \le P_*:&  \eps(P) &= 0~,\label{eps2}
\end{eqnarray}
where $\eps_s:=- ln(\beta)~D/(\chi_2-\chi_1)$ is a material constant
---  See Supplemental Material at [URL will be inserted by publisher] for details.

\begin{figure}[tb]
\includegraphics[width=1\linewidth,clip,viewport=0 -10 440 160]{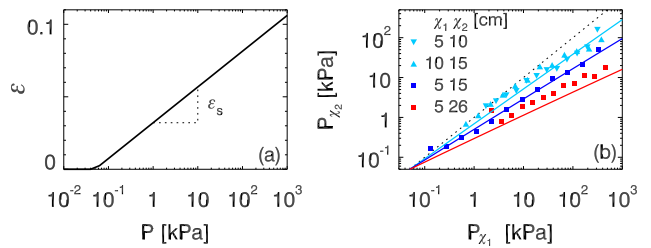}
\caption{(Color online) (a) Attenuation per contact $\eps(P)$. (b) Scatter plot of $P_1$
and $P_2$ for $P_0=1.4$ kPa at various locations $\chi_1$ and $\chi_2$, and corresponding fits
to our model.}
\end{figure}

A surprising consequence of this model is that it predicts that
the exponent $\beta$ should depend on the distance between $\chi_2$ and $\chi_1$.
To test this, we have performed several sets of experiments where we vary this
distance. Fig.~5b shows that the log slope relating $\Pone$ and $\Ptwo$ indeed increases for
larger propagation distance $\chi_2-\chi_1$ --- these trends are captured  by
our model without additional fit parameters.

{\em Local Elastic Motion and Energy Balance ---} As Fig.~5a illustrates, even for the strongest impacts we can access, the attenuation per particle is not more than 10\% --- it is likely smaller, as our model underestimates the number of contacts per unit length in a disordered media. Hence, we believe that the interactions in the leading edge of the shock wave are predominantly
elastic. Additional evidence for this
comes from a direct comparison between the contact forces and particle displacements; these are consistent with purely elastic (Hertzian) deformations (See Supplemental Material at [URL will be inserted by publisher] for details), showing that sliding and rearrangements do not play a dominating role in the leading edge of the shock.

This motivates
us to probe the balance between elastic and potential energies in our shock waves.
For the
conservative shocks studied in the numerical simulations \cite{gomez},
the kinetic and potential energies balance in the shock regime, whereas
away from the shock regime,
the kinetic energy tends to zero,
but the potential energy saturates at its lower bound $\sim
P_0^{5/3}$.

To probe this balance in our experimental data, we
present scatter plots of the peak velocities $u_1$
and $u_2$ vs the peak pressures $\Pone$ and $\Ptwo$ in Fig.~6.
The kinetic energy simply
scales $\sim u_m^2$ where $u_m$ is the maximum local velocity,
whereas the potential energy for Hertzian contacts
scales $\sim \delta_m^{5/2} \sim f_m^{5/3}$, where $\delta_m$ and
$f_m$ are the local maximum deformations and forces --- note that
the contact force $f_m$ should contain both the DC component
$\propto P_0$ and AC component $\propto {P_{\chi_{1,2}}}$.
Our data shows a
strikingly good qualitative agreement with the numerical data by
Gomez {\em et al.} (Fig.~3d of \cite{gomez}): it is fully consistent with a
$5/6$ power law in the strongly nonlinear regime
\cite{notesaturate}, and a pressure dependent deviation from this
law in the weakly nonlinear regime. Moreover, an estimate of
the kinetic and potential energies per particle based on the data shown in Fig.~6, shows that they are of similar order, just as in the simulations \cite{gomez}. We
finally  note that this balance is equally good
at $\chi_1$ and $\chi_2$, even though the energies in $\chi_2$ are
smaller than in $\chi_1$ due to attenuation.
All this suggest that the attenuation does not significantly upset the balance between
kinetic and potential energies.

\begin{figure}[tb]
\includegraphics[width=1\linewidth,clip,viewport=0 140 420 360]{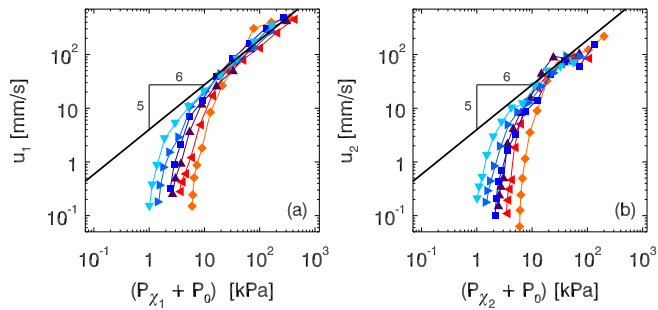}
\caption{(Color online) A comparison of the peak velocities and peak
pressures at (a) $\chi_1=5$ cm and (b) $\chi_2=15$ cm shows a striking similarity with the numerical data of Gomez {\em at.}.}
\end{figure}

{\em Discussion---} By varying the impact strength, we can tune our
waves from pressure dependent sound waves at low
impact, to pressure independent nonlinear shock waves at higher
impact, similar to what was predicted for dissipationless Hertzian particles \cite{gomez}. We find that propagating from one grain
to the next, a small amount of energy is dissipated, which
leads to
novel powerlaw scaling of the attenuation of the shocks amplitude
over several decades. Is this attenuation caused by scattering or dissipation? While we cannot rule
out scattering, we note that the overall decrease of the pressure and velocity profiles shown in Fig.~2 favor a dissipative picture; in addition, the width of the leading edge does not change very much under propagation. Notwithstanding this weak dissipation, the magnitude of displacements in the shock, and the balance of kinetic and potential energy, show that the physics in the leading edge of the shock is dominated by elastic interactions.

The shock waves we observe here are therefore qualitatively different from the slow granular densification waves known as ``plowing'' \cite{plow,cornstarch}. Plowing is associated with densification through highly dissipative rearrangements, and such densification fronts propagate
with velocities far below the sound speed \cite{plow}. Similarly slow and dissipative events may occur in the tail of our waves, but the point is that the leading edge of our shock waves propagate faster than the sound speed. They are also qualitatively different from the weakly nonlinear waves observed under continuous driving \cite{weakening,jia}.

Finally we note that there are two pressure scales that set the physics of granular shock waves in experiments--- the external pressure $P_0$ that sets the crossover from linear to shock waves, and the characteristic pressure $P^*$ above which attenuation sets in. In our experiment, $P_0 \gg P^*$, but it is conceivable that for
more elastic particles, or in microgravity, one can reach $P_0 \ll P^*$, in which case, virtually dissipation free granular shock waves could be produced.

\begin{acknowledgements}  We acknowledge discussions with J. Burton, L. Gomez, H. Jaeger, X. Jia,
and V. Vitelli, and funding from FOM, SHELL and NWO.
\end{acknowledgements}

\bibliographystyle{apsrev}

\begin{thebibliography}{36}
\bibitem{footnote1} The only counter example of a system retaining finite
elastic moduli near jamming we know of
are
frictional harmonic particles, which are not isostatic \cite{review1}.
\bibitem{shocknote}We use the term shock waves here in a loose sense,
meaning strongly nonlinear wave.
\bibitem{footnote2} For purely 2D Hertzian contacts, $f\sim
\delta^{\alpha}$ with $\alpha=1$, although 2D discs in experiments
have somewhat larger exponents $\alpha$ \cite{bobPRL}.
\bibitem{notesaturate} For the highest impacts, our accelerometers
saturate, making an accurate estimate of $u_m$ impossible --- such
points have been omitted in Fig.~4, which explains the somewhat
smaller range of $P$ in comparison to Fig.~3.
\end{thebibliography}

\newpage

{\bf Supplementary Material}

{\em Experimental Details --- } To detect the propagation of waves throughout our granular material, we employ several Br\"uel \& Kjaer
sensors. First, at each location $\chi_1$, $\chi_2$ we use pairs of accelerometers (Deltatron 4508 B002, 1 V/g and 4508 B001, 10 mV/g, resonance at 26 kHz,
dimensions 1x1x1.6 cm, weight 4.8 g). Together these cover the wide range of
accelerations that we encounter (from
less than 0.1 m/s$^2$ to larger than 7000 m/s$^2$), although they saturate for some of the very strong impacts, which we conclude to lead to accelerations exceeding 700 g!
The accelerometers allow us, by integration, to measure the local
particles velocities and displacements.

We also use force sensors (Type 8230, 112.5 mv/N at $\chi_1$ and 82301-001,
22.05mV/N at $\chi_2$, resonance at 75 kHz, cylindrical shape 19 mm long, circular contact area 1.33 cm$^2$, weight 30 g). These do not saturate for our experiments and
we focus on the signals from these sensors in most of our analysis. The sensors
are much  stiffer (2kN/$\mu$m) than the typical stiffnesses of the Hertzian contacts we encounter, and are immersed in the grains, thus acting as  pressure sensors. They have no DC response, so that they only are sensitive to the deviation from the confining pressure. To convert force to pressure, we divide by their circular contact area.

In our experiments, we vary the pressure $P_0$ at the depth of the sensors from 0.9 to 6 kPa by varying the load on the topplate. For each $P_0$, we perform several series of experiments, each consisting of hundreds of impact experiments. We vary the impact strength (characterized by the peak pressure) over four decades: lightly tapping the resting pendulum generates the weakest impacts we can detect, while swinging the pendulum with full force from its
maximum height generates the strongest impacts --- only at the
lowest confining pressures and at the strongest impacts, the
piston penetrated the packing significantly, limiting the number
of experiments in this regime. Before each series of experiments the container was freshly filled and then tapped to stabilize the packing, while after each series the container was emptied and the correct placing of the detectors
was verified. No systematic dependence on filling procedure or ''age'' of the sample was detected. For each impact, the signals from
all six sensors were recorded at 250 kHz, and  data sets ranging
from 300 point before to 2000 points after impact were stored. To accurately
determine the timing where the pressures reach their 50\% peak magnitude, we
perform linear fits to the 20\% - 80\% leading slope of the force signals.

\begin{figure*}[tb]
\includegraphics[clip,viewport=10 60 480 280,width=1\linewidth]{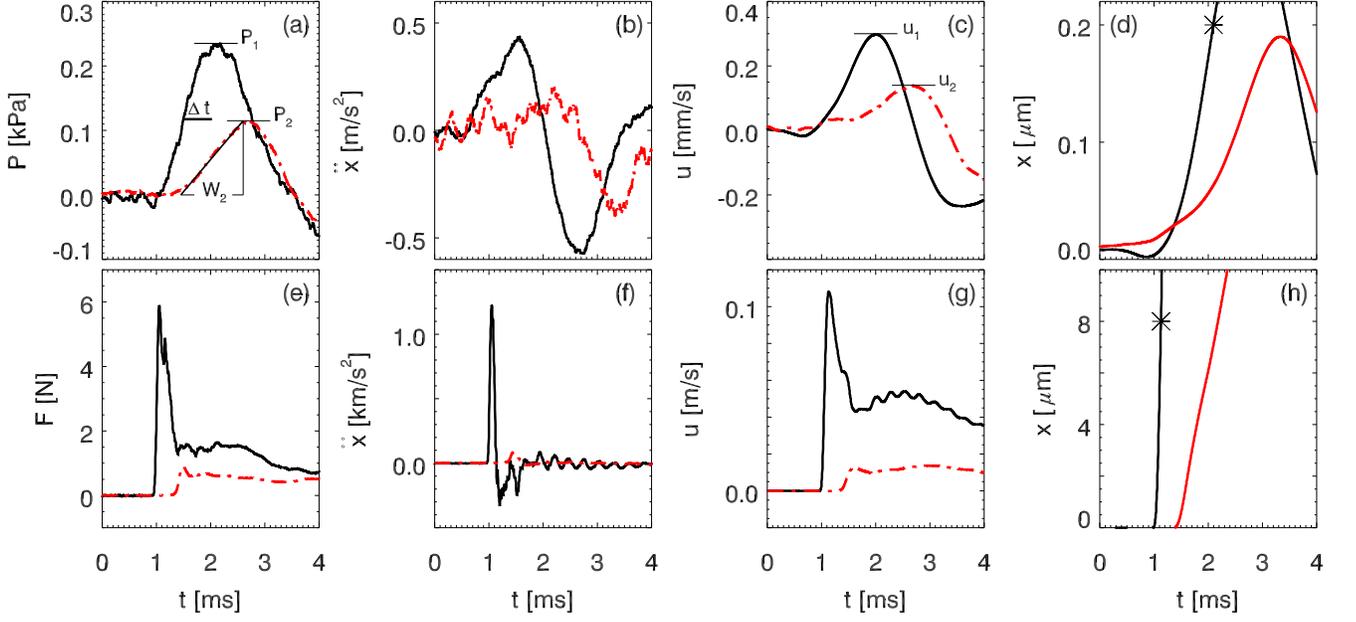}
\caption{(Color online) Typical signals for low (a-d) and high
(e-h) impact amplitude, for $P_0=0.9$ kPa, at $\chi_1=5$ cm (black)
and $\chi_2=15$ cm (red/dashed). (a) Local pressure $P(t)$, and
definition of peak pressures $P_1$, $P_2$, slope (only for $W_2$),
and $\Delta t$ as delay between midpoints of slopes.  (b) Local
acceleration $\ddot{x}(t)$. (c) Local velocity $u(t):=\int_0^t
d\tau \ddot{x}(\tau)$. (d) Local displacement $x:=\int_0^t d\tau
u(\tau)$ - the star indicates the peak time of the pressure. (e-h)
Same signals, now for much larger impact. Note: the data in panel e is identical
to the data shown in Fig.~2a in the main text, but here we take a linear scale and the raw contact force instead of pressure.}\label{figapp1}
\end{figure*}

{\em Wave profiles --- } Our sensors allow us to extract the time evolution of the
local grain pressure,
acceleration, velocity and displacement. Fig.~\ref{figapp1} shows representative
examples of these profiles, both for a weak and a strong impact, qualitatively illustrating important aspects of the phenomenology of the waves we observe.

These signals allow us to extract typical scales characterizing the wave profiles.
The first striking observation is that even for strong impacts, the displacements
are rather small, and consistent with Hertzian deformations. Assuming that approximately
10 particles make contact with the sensor, we find that for the string shock wave shown
in Fig.~\ref{figapp1}e-h, the peak force at $\chi_1$ is of order 0.6 N
(and is reached at time $t^*\approx0.05$ ms).
Using Hertz law, which states that the
contact force $F$ varies with particle deformation $\delta$ as
$F=\frac{4}{3}E^* R^{1/2} \delta^{3/2}$ \cite{Johnson},
and taking $E^* = 50$ Gpa (typical value for glass), we estimate that $\delta \approx 0.35 ~\mu$m. The location
$\chi_1$ is 12 particle diameters $\approx$ 24 contact zones away, so
the cumulative motion $x$ at $\chi_1$, assuming a fairly flat pressure
profile, is predicted to be of order $8 \mu$m, which is close to
the measured displacement at $t^*$ ($8 \mu$ m, see Fig.~\ref{figapp1}h). Such estimates are also reasonably good for other nonlinear waves, which suggests that in the leading edge of the wave, the deformations are predominantly elastic, and grain sliding and rearrangements have not arisen yet (these do arise in the long tail of the wave \cite{weakening,jia}). The elastic nature of the contacts also explains why the pressure
and velocity signals look qualitatively similar --- local forces
are proportional to the pressure, while local deformations
$\delta$ of the grains are given by the {\em gradient} of $x$,
which, assuming that the pulse moves with constant speed $v_s$,
is proportional to the local grain velocity $u_s$.

{\em Model for Attenuation ---}
In our simple model for attenuation, we ignore disorder and thus
imagine the signal to propagate over a linear chain of beads, where each bead attenuates the signal by a factor $(1-\varepsilon)$ --- see Fig.~8a. We take $D$ as our length scale ($\tilde{x}:= x/D$), define $S:=(\chi_2-\chi_1)/D$, and in Eq.~(\ref{disc}-\ref{lastdimless}) below drop the tildes for simplicity, so that
\begin{equation}\label{disc}
P^{i+1} = \left(1-\varepsilon(P^i)\right) P^i~,
\end{equation}
and after taking the continuum limit we obtain:
\begin{equation}\label{ODE}
\frac{dP}{dx}=-\eps(P) P~.
\end{equation}
To infer  $\eps({P})$ from
the observed relation between $P_2$ and $P_1$ (Eq.~\ref{pwrlw1}-\ref{pwrlw2})
is far from trivial, as $P_2(P_1)$ follows from integrating Eq.~(\ref{ODE}) over a finite distance $S$.

To determine $\eps(P)$ we note that solutions
of Eq.~(\ref{ODE}), $P(x)$, can all be derived from a single mastercurve ${\cal P}(\zeta)$: $P(x)={\cal P}(x-dx)$, where the shift $dx$ is adjusted to satisfy the initial condition of Eq.~({\ref{ODE}) --- see Fig.~8a. Hence, we obtain the following relation between ${\cal P}$, $P_1$ and $P_2$:
\begin{equation}
{\cal P}^{-1}(P_2)={\cal P}^{-1}(P_1) + S~.
\end{equation}
Focussing on the power law relation between $P_2$ and $P_1$ for $P\!<\!1$,
we need to determine ${\cal P}^{-1}$ that satisfies:
${\cal P}^{-1}(P_1^\beta) = {\cal P}^{-1}(P_1) +S~$,
which suggest that ${\cal P}^{-1}$ should be a double log function --- so as to turn the exponent $\beta$ into the additive term $S$. After some algebra, we obtain that ${\cal P}(\zeta)$ is a double exponential function:
\begin{eqnarray}
P^{-1}(\zeta) &=& \frac{S}{ln(\beta)} ln \left( ln (\zeta)\right) \Rightarrow\\
P(\zeta) &=& \exp(\exp(\frac{ln(\beta)}{S}\zeta))~,\label{lastdimless}
\end{eqnarray}
where we stress that ${\cal P}$ asymptotes to $P^*$, indicating the absence
of dissipation when $P \rightarrow P^*$  (Fig.~8b).

\begin{figure}[tb]
\includegraphics[width=1\linewidth,clip,viewport=0 160 440 337]{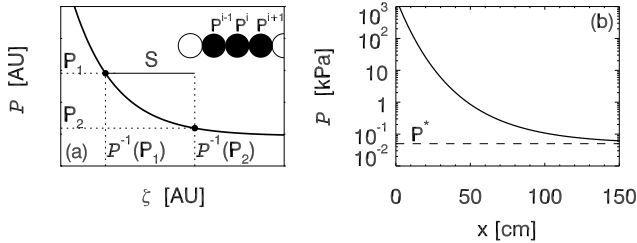}
\caption{(Color online) (a) Illustration of particle model and relation between ${\cal P}(\zeta)$, $P_1$, $P_2$ and $S$. (b) Actual solution for
${\cal P}(x)$.}
\end{figure}

This solution $P(\zeta)$ satisfies Eq.~\ref{ODE} when $\eps(P)$ has a logarithmic form, which in dimensional quantities is given by Eqs.~(\ref{eps1}) and (\ref{eps2}).

{\em Potential and Elastic Energies ---} The peak kinetic energy $U_k$ per particle
measured at $\chi_j$
equals the $(1/2)m u_j^2$, where for average grains ($r\approx 2$ mm),
$m \approx \rho ~ 4/3 \pi r^3 \approx 8.7~10^{-5}$ kg. The peak potential energy $U_p$ per contact equals $\int_0^{\delta} f(x) dx$, where $\delta$ is the maximum deformation and $f$ the contact force, which we assume to be Hertzian:
$f(\delta) =  K \delta^{3/2}$, where $K=4/3 ~ E^* r^{1/2}$.
Hence, $U_p= 2/5 ~ K \delta^{5/2} = 2/5 ~ K^{-2/3} f^{5/3}$.
Assuming that 10 particles
are in contact with the force sensor of area $A = 1.33~10^{-4}$ m$^2$, and taking
into account that the total pressure is the sum of $P_0$ and
$P_{\chi_{i}}$, we can estimate the individual contact forces as $f=A/10 ~ (P_0+P_{\chi_{i}})$.

Now assuming that the peak potential
energy of $z$ contacts equals the peak kinetic energy, we arrive at
\begin{eqnarray}
1/2 ~ m u_i^2 = 2/5 ~ z K^{-2/5} f^{5/3} \Rightarrow \\
u_i = \sqrt{\frac{z}{m}\frac{4}{5}K^{-2/5}(\frac{A}{10})^{5/3}} ~(P_0 + P_{\chi_{i}})^{5/6}~.
\end{eqnarray}
The straight lines in Fig.~6 correspond to $z=5$, which is a reasonable number
for frictional particles in 3D, which, in the limit of hard frictional spheres have $4<z<6$.

\end{document}